\documentclass[11pt]{article}

\usepackage{amssymb,amsfonts,psfig,epsfig,axodraw}
\newcommand{\bea}{\begin{eqnarray}}
\newcommand{\beq}{\begin{equation}}
\newcommand{\eea}{\end{eqnarray}}
\newcommand{\eeq}{\end{equation}}
\newcommand{\nn}{\nonumber}
\newcommand{\Frac}[2]{\frac{\displaystyle{#1}}{\displaystyle{#2}}}
\newcommand{\lsim}{\raise0.3ex\hbox{$\;<$\kern-0.75em\raise-1.1ex\hbox{$\sim\;$}}}
\newcommand{\gsim}{\raise0.3ex\hbox{$\;>$\kern-0.75em\raise-1.1ex\hbox{$\sim\;$}}}

\begin{document} 

\pagestyle{empty} 
\begin{flushright}\end{flushright}  

\vskip 1.5 cm    
\def\thefootnote{\fnsymbol{footnote}}
\begin{center}  
{\large \bf F-flatness and universality in broken Supergravity 
\protect\footnote{Proceedings
of SUSY02, 10th International Conference on {\it Supersymmetry and Unification of Fundamental
Interactions}, 17-23/06/2002, DESY, Hamburg, Germany.}}     
\end{center}  
\vspace*{5mm} 
\centerline{\bf Oscar Vives}
\vskip 0.2 cm
\centerline{\em Dep. of Physics, 
Theoretical Physics,}
\centerline{\em U. of Oxford, Oxford, OX1 3NP, UK}
\vskip .5 cm   
\centerline{\bf Abstract}  
We show that in a broken supergravity theory any field
that acquires a vacuum expectation value obtains at the same time a non 
vanishing F-term whose natural size is given by 
$\langle \theta \rangle m_{3/2}$. These F-terms contribute unsuppressed
to the trilinear couplings and have observable effects in the low energy 
phenomenology. We show that $\mu \to e \gamma$ decays can receive
large contributions from this source at the level of current experimental
bounds. Therefore this decay may provide the first clue of the structure
of the soft breaking sector with the proposed experiments in the near future.
\vspace*{.5cm} 
\pagestyle{plain}


\def\thefootnote{\arabic{footnote}}
\setcounter{footnote}{0}

\section{Motivations}

Our knowledge of the Supersymmetry breaking sector in the  Minimal 
Supersymmetric extension of the Standard Model (MSSM) is still very
small due to the limited experimental information available.
Only after the discovery of SUSY particles and the measurement of the
Supersymmetric spectrum we will be able to explore in detail  
and improve our understanding on this fundamental piece of the MSSM.
Nevertheless, we have already today a lot of useful information on
this sector from experiments looking for indirect effects of SUSY
particles in low-energy experiments \cite{Annrev}.
In fact, it was readily realized at the beginning of the SUSY phenomenology
era that large contributions to Flavour Changing Neutral Currents (FCNC)
and $CP$ violation phenomena were expected in Supersymmetric theories with
a generic soft breaking sector. The absence of these effects much beyond
the level most theorists considered reasonable became to be known as the
SUSY flavour and $CP$ problems.    
These problems are closely related to the flavour structure of the soft
breaking sector and therefore to the source of flavour itself. 
In fact, the main solutions to these problems are either exact
universality of the soft breaking terms or alignment to the Yukawa
matrices.  However, as we will show here, in any model where we intend
to explain at the same time the strongly hierarchical structure of the
Yukawa matrices, these two solutions are not perfect and the soft
breaking terms are necessarily nonuniversal and not aligned with the
Yukawa couplings. This is due to the fact that in a theory with broken local 
Supersymmetry, i.e. Supergravity, any field that acquires a vacuum 
expectation value (vev) obtains at the same time a non-vanishing F-term
\cite{rossvives,abelservant}.

To prove this we consider a generic supergravity scenario specified in 
terms of the superpotential $W$ and the K\"ahler potential,
$K = \hat K(\chi, \chi^*) + \sum_i K^i_i(\chi, \chi^*)  
|\varphi_i|^2$.
Where $K$ is a real function of the chiral superfields and $\chi$ are 
general hidden sector fields. In these conditions, 
the F-term contribution to the scalar potential is given by, 
\bea
V = e^{K} \left[ \sum_i (K^{-1})^j_i \tilde F^i \tilde F_j - 3 |W|^2 \right]
\label{scalpot}
\eea
with $\tilde F^i = \partial W /\partial \phi_i + K^i_i 
\phi^{i *} W$ and $\tilde F_i = \partial W/\partial \phi^{i *} 
+ K^i_i \phi_{i} W^*$ related to the correctly normalized supergravity 
F-terms as $F^i = e^{K/2} \tilde F^i$ and $(K^{-1})^i_j$ is the inverse 
of the K\"ahler metric.
We study a situation of broken supergravity where 
$m_{3/2}= e^K/2 \langle W \rangle \neq 0$ and we have 
a non-vanishing vev for a certain field $\theta$ after minimization of the 
scalar potential. In particular we are interested in the case of
fieds that obtain small vevs in units of the Plank mass and generate 
the hierarchy in the Yukawa couplings. So, we take this field as a singlet 
under the SM gauge symmetries but charged under a new flavour symmetry 
that controls the Yukawa structure. 
The natural scale of the F-terms with $m_{3/2} \neq 0$ is given by
$m_{3/2}\langle \theta \rangle$ as this is the supergravity correction
to the globaly sypersymmetric F-term. However this natural size could be
reduced by some cancellation from other terms. 
In the case of a continuous flavour symmetry controling the Yukawa hierarchy
and when only the field $\theta $ acquires a vev,
we have $\partial W /\partial \theta = 0$ at the minimum and the vev of 
$\theta$ will be driven by some scale dependent soft masses in analogy to the 
radiative symmetry breaking in the MSSM \cite{radiative}. Hence it is
strait-forward from the definition that $F^\theta = m_{3/2}\langle \theta
\rangle$ as expected.

On the other hand in the presence of a discrete symmetry or a continuous 
symmetry with several fields $\phi _{i}$ which
acquire vevs, the situation is 
more involved. The superpotential is of the kind, $W~=~a~+~b~\theta ^{p}
\phi ^{q}$ and the relevant part of the scalar potential is (after 
absorbing the K\"ahler metric in a redefinition of the fields),
\bea
V_{\theta \phi }~~=~~\left|p~b~\theta ^{p-1}\phi ^{q}+\theta ^{\ast
}m_{3/2}\right|^{2}+ \left| q~b~\theta ^{p}\phi ^{q-1}+\phi ^{\ast
}m_{3/2}\right|^{2}-
3\left|m_{3/2}+b~\theta ^{p}\phi ^{q}\right|^{2}+\nn
\\
m_{r}^{2}(\theta)\left|~ \theta~\right| ^{2} + A_r(\theta)~ b~(p+q-3) 
\left(\theta^{p}\phi ^{q}+ \theta ^{p\ast}\phi ^{q\ast}\right)
\label{scalar2}
\eea
where we allowed the generation of D-terms at radiative order, 
$m_{\theta }^{2}(\theta ),$ in order to generate a minimum with 
$\theta \neq 0$.  D-flatness aligns the $\theta $ and $\phi$ vevs and we
get $|\theta|^2/|\chi|^2 = p/q$
$\theta~ \partial V/\partial \theta~ -~\theta^*~ \partial
V/\partial \theta^*~=0$ implies that 
$\mbox{Im}~ \theta^p \chi^q = 0$ and for simplicity ($p$ or $q$ odd), 
we choose them as real.
Using these relations and after some algebra, we obtain in terms of a
new variable  $X\equiv \Frac{b}{m_{3/2}}~ p~
\left(\Frac{q}{p}\right)^{q/2}~ \theta^{p+q-2}$,
\bea
\Frac{\partial V}{\partial \theta}~=~ m_{3/2}~ \theta~
\left[X^2(p+q-1) + X~(1-\alpha)~(p+q-3)+ 1 - \beta
\right] =0 
\label{change}
\eea 
where $\alpha\equiv\Frac{A(\theta)}{m_{3/2}}$ and
$\beta\equiv\Frac{m_1 (\theta)^2}{m_{3/2}^2}\Frac{p}{p+q}$. The
solutions with nonvanishing $\theta$ vev correspond to the roots in
the quadratic equation in $X$. In terms of $X$ the F terms are now,
\bea
 F^\theta = ~m_{3/2}~ \langle \theta \rangle ~(1
 + X) \equiv \gamma~ \langle \theta \rangle~ m_{3/2}
\label{Fterm}
\eea
and similarly for $F^\chi$. Clearly there is no reason to expect a
conspiracy 
between radiative terms and the Superpotential structure to obtain 
$X\simeq -1$ and so $\gamma ={\cal{O}}(1)$. Thus, terms
involving $\theta$ also contribute to the soft SUSY breaking terms.

At this point, it is clear that any field $\theta$ that obtains a vev in
the presence of broken SUSY, acquires simultaneously a non-vanishing
F-term of order $\langle \theta \rangle m_{3/2}$ \footnote{As shown in 
\cite{rossvives} it is still possible to get a further reduction if the 
familon field has a vev in the SUSY limit. However, to obtain this vev 
requires a significantly more complicate theory and this is the 
correct order of magnitude for the F-terms in most simple models.}. 
Thus, they also contribute to the SUSY soft breaking terms. However,
these contributions are in principle small, suppresed by the small vev
of $\theta$, when compared with the dominant contributions of order 
$m_{3/2}$. Therefore, the main question now is whether these 
additional contributions can have any sizeable effect in the low energy
physics. So we analyze the effective theory below the scale of
flavour symmetry breaking with a $W$,
\bea
W = W^{hid} (\chi_k) + \left(\Frac{\theta}{M}\right)^{\alpha_{ij}} H_a 
Q_{L i} q^c_{R j} + \dots
\label{SW}
\eea
where the hierarchical structure of the Yukawa couplings is generated through 
effective operators in terms of $\theta$ vevs suppressed by 
a heavy mediator scale which can be $M_{Pl}$. 

Then, from Eq.~(\ref{scalpot}) we get the soft breaking terms in the 
observable sector after SUSY breaking. 
At the minimum we obtain the trilinear terms as \cite{soni},
\bea
A_{i j} \hat Y^{i j} = 
F^{\chi_k} \hat K_{\chi_k} Y^{i j} +\alpha_{ij}\ \Frac{e^{K/2}}{M} 
\left(\Frac{\theta}{M}\right)^{\alpha_{ij} -1 } m_{3/2} \theta 
\label{nunitri}
\eea
with $\hat K_{\chi_k}= \partial \hat K / \partial \chi_k$ and 
$Y^{i j}= e^{K/2} (\theta/M)^{\alpha_{ij}}$. 
The presence of the $\alpha_{ij}$ in the right hand side is due to
the dependence of the effective Yukawa couplings on $\theta$.  
Still the smallness of $\theta$ does not affect the trilinear
couplings because is reabsorved in the Yukawa coupling itself,
\bea
m_{3/2}\  \langle \theta \rangle\  \Frac{\partial Y^{i j}}{\partial \theta} = 
\alpha(i,j)\  m_{3/2}\  Y^{i j} .
\eea
Therefore, in any framework explaining the hierarchy in the Yukawa textures 
through nonrenormalizable operators, the trilinear
couplings are necessarily nonuniversal as in Eq.~(\ref{nunitri}). 
In a similar way, we can also expect non-renormalizable contributions to 
the K\"ahler potential of the kind $(\theta \theta^*/ M^2)^{\alpha(i,j)}$.
Nevertheless these contributions appear only at order $2 \alpha(i,j)$ in
$\theta/ M$ with respect to the dominant term ${\cal O} (1)$.
So, in the following we concentrate in the nonuniversal trilinear couplings.

Next, we must check whether this breaking of universality does not
contradict any of the very stringent bounds from low energy phenomenology. 
Unfortunately we do not have still a unique and complete
theory of flavour that provides the full field deppendence of the low
energy effective Yukawa couplings \cite{flavourth}. Still, the analysis 
of the fermion masses and mixing angles seems to point to some definite 
textures for the Yukawa matrices \cite{liliana},
\begin{equation}
\frac{M}{m_{3}}=\left( 
\begin{array}{ccc}
0 & b \epsilon^{3} & c\epsilon^{3} \\ 
b^\prime \epsilon ^{3}& d \epsilon ^{2} & a \epsilon^{2} \\ 
f \epsilon ^{m} &  g \epsilon ^{n} & 1
\end{array}
\right),  \label{texture}
\end{equation}
with $\epsilon_d=\sqrt{m_s/m_b}= 0.15$ and $\epsilon_u=\sqrt{m_c/m_t}=0.05$.
at the unification scale, and $a,b, b^\prime,c,d,g,f$ coefficients 
${\cal{O}}(1)$ and complex in principle. In this 
texture there are still two undetermined elements, $(3,1)$ and $(3,2)$, 
that would account for the unmeasured right-handed quark mixings 
although fermion masses require $m \geq 1$. Next, we assume that the 
hierarchy in the 
Yukawa matrices is generated by different powers in the vevs of scalar 
fields, $\epsilon_a= \langle \theta_a \rangle / M$, and  we 
can immediately calculate the nonuniversality in the trilinear terms, 
$(Y^A)_{i j} \equiv Y_{ij} A_{ij} $, 
\bea
(Y^A)_{i j} =  A_0 Y_{ij} + m_{3/2}\ 
Y_{33} \left( 
\begin{array}{ccc}
0 & 3 b \epsilon^{3} & 3 c\epsilon^{3} \\ 
3 b \epsilon ^{3}& 2 d \epsilon ^{2} & 2 a 
\epsilon^{2} \\ 
f m \epsilon^{m} & g n \epsilon ^{n} & 0
\end{array}
\right)
\label{trisymm}
\eea
with $A_0 = F^{\chi_k} \hat K_{\chi_k}$. The Yukawa texture in 
Eq.~(\ref{texture}) is diagonalized by superfield rotations in the so-called
SCKM basis, $\tilde Y = V_L^\dagger\cdot Y\cdot V_R$.  
However, in this basis large off-diagonal terms necessarily remain in the
trilinear couplings, 
$\tilde Y^A = V_L^\dagger\cdot Y^A\cdot V_R$.  
The phenomenologically relevant flavour off-diagonal entries for 
FCNC contributions in the basis of diagonal Yukawa matrices are,
\bea 
(\tilde Y^A)_{3 2} \simeq Y_{3 3}\ m_{3/2}\ g\ n\ \epsilon^n + \dots~~~~~~~~ 
\\
(\tilde Y^A)_{2 1} \simeq Y_{3 3}\ m_{3/2}\ \epsilon^3 (b^\prime + a\ ( 
\Frac{b^\prime}{d}\ g\ n\ \epsilon^{n} - f\ m\ \epsilon^{m-1})) \nn
\eea
We take these matrices as the boundary conditions at a high scale, typically 
$M_{GUT}$. Then we must use the MSSM Renormalization Group Equations
(RGE) \cite{RGE} to obtain the corresponding matrices at the electroweak
scale. The main effect
in this RGE evolution is a large flavour universal gaugino contribution
to the diagonal elements in the sfermion mass matrices (see for instance
Tables I and IV in \cite{wien}). In this minimal supergravity scheme we 
take gaugino masses as $m_{1/2}= \sqrt{3} m_{3/2}$ and sfermion masses
$m_0^2=m_{3/2}^2$. So, the average squark and slepton masses,
\bea&
m^2_{\tilde{q}} \simeq 6 \cdot m_{1/2}^2 + m_0^2 \simeq 19\ m_{3/2}^2,
~~~~~~~m^2_{\tilde{l}} \simeq 1.5 \cdot m_{1/2}^2 + m_0^2 \simeq 5.5\ m_{3/2}^2&
\label{average}
\eea
The RG evolution of the trilinear terms is also similarly dominated 
by gluino contributions and the third generation Yukawa couplings.
However, the offdiagonal elements in the down and 
lepton trilinear matrices are basically unchanged for $\tan \beta \leq 30$ 
\cite{wien}. From here we can obtain the full trilinear couplings and 
compare with the experimental observables at low energies. 
The so-called Mass Insertions (MI) formalism is very useful in this 
framework. The left--right MI are defined in the 
SCKM basis as $(\delta_{LR})_{ij} = (m^2_{LR})_{ij}/m^2_{\tilde{f}}$, 
with $m^2_{\tilde{f}}$ the average sfermion mass.
\begin{table}
 \begin{center}
 \begin{tabular}{||c|c|c|c|c||}  \hline \hline
  $x$ &
 ${\sqrt{\left|\mbox{Im}  \left(\delta^{d}_{LR} \right)_{12}^{2}
\right|} }$ & $\sqrt{\left|\mbox{Re}  \left(\delta^{d}_{13}
 \right)_{LR}^{2}\right|} $ &  ${\left|\left(\delta^{l}_{LR} \right)_{12}
\right|}$ &
 ${\left|\left(\delta^{l}_{LR} \right)_{23}
\right|}$ \\
 \hline
 $
   0.3
 $ &
$
1.1\times 10^{-5}
 $ & $
1.3\times 10^{-2}
 $ & $
6.9\times 10^{-7}
 $ & $
8.7\times 10^{-3}
 $ \\
 $
   1.0
 $ &
 $
2.0\times 10^{-5}
 $ & $
1.6\times 10^{-2}
 $ & $
8.4\times 10^{-7}
 $ & $
1.0\times 10^{-2}
 $  \\
 $
   4.0
 $ &
 $
6.3\times 10^{-5}
 $ & $
3.0\times 10^{-2}
 $ & $
1.9\times 10^{-6}
 $ & $
2.3\times 10^{-2}
 $ \\ \hline \hline
 \end{tabular}
\caption{MI bounds from $\varepsilon^\prime/\varepsilon$, $b \to s
\gamma$,  $\mu \to e \gamma$ and $\tau \to \mu \gamma$ for an average 
squark mass of $500\mbox{ GeV}$ and different values of 
$x=m_{\tilde g}^2/m_{\tilde q}^2$ or an
average slepton mass of $100\mbox{ GeV}$ and different values of 
$x=m_{\tilde \gamma}^2/m_{\tilde l}^2$.
These bounds scale as $(m_{\tilde f} (\mbox{GeV}) / 500 (100))^2$ for 
different average sfermion masses}
\label{tab:MI1}
 \end{center}
 \end{table}
We can estimate the value of $(\delta_{LR})_{21}$ as,
\bea
&\left(\delta^d_{LR}\right)_{2 1} \simeq \Frac{m_b\,\epsilon_d^3}{19\,
 m_{3/2}} 
(b^\prime\,+ a\,\Frac{b^\prime}{d}\,g\,n\,\epsilon_d^{n} 
-  a\,f\,m\,\epsilon_d^{m-1}) \simeq\nn \\& (b^\prime\,+ 
a\,\Frac{b^\prime}{d}\,g\,n\,\epsilon_d^{n} - a\,f\,m\,
\epsilon_d^{m-1})\,7.5\times 10^{-6}
\label{eps'}
\eea 
using $m_{\tilde q} \simeq 500$ GeV corresponding to $m_{3/2} \simeq 120$ 
GeV and $\epsilon_d\simeq 0.15$ \cite{liliana}. 
We can compare our estimate for the mass insertion with the phenomenological
bounds in Table \ref{tab:MI1} with $x=m_{\tilde g}^2/m_{\tilde q}^2 \simeq 1$.
Even allowing a phase ${\cal{O}}(1)$, necessary to contribute to 
$\varepsilon^\prime/\varepsilon$, we can see the bound requires 
only $m \geq 1$ which is already required by fermion masses. Still, we can 
see that in the presence of a phase, 
$\varepsilon^\prime/\varepsilon$ naturally receives a sizeable contribution 
from the $b^\prime$ term \cite{murayama}. 

The MI corresponding to the $b \to s \gamma$ decay are, 
\bea
\left(\delta^d_{LR}\right)_{3 2} \simeq \Frac{m_{3/2}\,m_b\,g\,n\,
\epsilon_d^n}{19\,m_{3/2}^2} \simeq 2.2 \times 10^{-3}\,g\,n\,\epsilon_d^n 
\eea
again with $m_{3/2}\simeq 120$ GeV. This estimate is already of the same 
order of the phenomenological bound for any $n$ and  
we do not get any new constraint on $n$.

The situation is more interesting in the leptonic sector. Here, the 
photino contribution is indeed the dominant one for LR mass insertions. 
In Table \ref{tab:MI1} , we show the rescaled 
bounds from Ref. \cite{gabbiani} for the present limits on the branching 
ratio.
In this case, it seems reasonable to expect some kind of 
lepton-quark Yukawa unification and we assume that  
the charged lepton Yukawa texture at $M_{GUT}$ shares the same structure as 
the down Yukawa, except for the Georgi-Jarskog factors of 3 in (2,2), (2,3) 
and (3,2) entries. Therefore we obtain,
\bea
&\left(\delta^e_{LR}\right)_{1 2} \simeq \Frac{m_\tau\,\epsilon_d^3}
{5.5\,m_{3/2}} (b^\prime + 9\,a\,\Frac{b^\prime}{d}\,g\,n\,
\epsilon_d^{n} -3\,a\,f\,m\,\epsilon_d^{m-1}) \simeq& \nn \\ 
&(b^\prime\,+ 9\,a\,\Frac{b^\prime}{d}\,g\,n\,\epsilon_d^{n} -
3\,a\,f\,m\,\epsilon_d^{m-1})\,8.7 \times 10^{-6}&
\label{mueg}
\eea
where we take $m_{3/2} \simeq 120$ GeV corresponding to 
$m_{\tilde l}= 280$ GeV. Notice that, at least, we have an unavoidable 
contribution from the $b^\prime$ entry, 
$\left(\delta^e_{LR}\right)_{1 2} \simeq b^\prime\,8.7 \times 10^{-6}$. 
This must be compared with
a bound, $\left(\delta^e_{LR}\right)_{1 2} \leq 7 \times 10^{-7}\  
(280/100)^2 = 5.5 \times 10^{-6}$. 
Clearly the estimate in Eq.~(\ref{mueg}) is still too large when compared 
with this experimental bound. Therefore this bound necessarily requires a 
larger value of the slepton mass. For an average slepton mass of 320 GeV 
($m_{3/2}= 136$ GeV, $m_{\tilde q}= 600$ GeV) our estimate would be just 
below the MI bound. Again, there are no new constraints on the value 
of $m$ and $n$. 
It is well worth to recall the very important results we get from
this observable: Assuming a quark-lepton unification at $M_{GUT}$,
nonuniversality in the trilinear terms predicts a large 
$\mu\to e \gamma$ branching ratio even beyond the values expected from other
sources as SUSY seesaw \cite{SUSYseesaw}. This points out this 
decay as the most sensitive probe of SUSY and the soft breaking sector 
in the near future.

Another interesting constraint in this scenario is
provided by Electric Dipole Moment (EDM) bounds. Even
in the most conservative case, where all soft SUSY breaking parameters
and $\mu$ are real, we know that the Yukawa matrices contain phases
${\cal{O}} (1)$. If the trilinear terms are nonuniversal,
these phases are not completely removed from the diagonal elements of 
$Y^A$ in the SCKM basis and hence can give rise to large EDMs \cite{abel}.  
However, it is possible to prove that the phase in the trilinear terms 
will be exactly zero at leading order in $\theta$ for any diagonal element. 
To see this we must take into account that the eigenvalues and mixing
matrices of the yukawa matrix deppend on $\theta$, $Y(\theta) = V_L^\dagger 
(\theta)\, D (\theta) \, V_R (\theta)$. The contribution to the 
trilinear terms proportional to $\theta\  \partial Y/\partial \theta$ is, 
\bea 
V_L \theta \Frac{\partial Y}{\partial \theta} V_R^\dagger =
V_{L}\Frac{\theta \partial V_{L}^\dagger}{\partial \theta} D +
\Frac{\theta \partial D }{\partial \theta} +
D \Frac{\theta \partial V_{R}}{\partial \theta} V_{R}^\dagger
\label{deriv}
\eea
In this expression the dominant contribution in $\theta$ is controled by the
lowest power in $D_{ii}$. Clearly $\theta\ \partial V/\partial \theta$
always adds at least a power of $\theta$ and therefore the first and
third terms in the above equation can only contribute to subdominant terms in
the $\theta$ expansion for the diagonal elements. Hence the dominant term 
in a diagonal element is provided by the second term and is exactly 
proportional to the leading $\theta$ term in $Y_{ii}$ with a coefficient 
equal to its power in $\theta$.
This implies that any observable phase in the diagonal elements will only 
appear at higher orders, for instance if $n \geq 1$ or $m \geq 2$ or with
higher order contributions to entries of the Yukawa matrix.
With real  $\mu$ and soft breaking, EDMs are generated by,
\bea
\mbox{Im}\left(\delta^{q,l}_{LR}\right)_{1 1} \simeq 
\Frac{m_1}{R_{q,l}\,m_{3/2}} \left( \epsilon^{n}\,n\, +
\epsilon^{m-1}\,(m-1)\,\right)
\eea 
where we use $m_1 = m_3\,\epsilon^4 b b^\prime/d$ and then take
all unknown coefficients ${\cal{O}}(1)$.
The coefficients $R_{q}=19$ and $R_l=5.5$ take care of the RGE effects 
in the eigenvalues as before. So, taking into account that 
$\epsilon_d=0.15$, $\epsilon_u=0.05$ and $m_d \simeq 10$ 
MeV, $m_u \simeq 5$ MeV and $m_e = 0.5$ MeV and assuming phases 
${\cal{O}}(1)$, we get,
\bea
\mbox{Im}\left(\delta^{d}_{LR}\right)_{1 1} \simeq \left( \epsilon^{n}_d\,n\,
 + \epsilon^{m-1}_d\,(m-1)\,\right) \,3.9 \times 10^{-6} \nn \\
\mbox{Im}\left(\delta^{u}_{LR}\right)_{1 1} \simeq \left( \epsilon^{n}_u\,n\,
 + \epsilon_u^{m-1}\,(m-1)\,\right) \,1.9 \times 10^{-6} \nn \\
\mbox{Im}\left(\delta^{e}_{LR}\right)_{1 1} \simeq \left( \epsilon^{n}_d\,n\,
 + \epsilon^{m-1}_d\,(m-1)\,\right) \,6.7 \times 10^{-7} 
\eea
\begin{table}
\begin{center}
\begin{tabular}{|c||c|c|c|c|}
\hline
  $x$   & $\vert {\rm Im}(\delta_{11}^{d})_{LR}\vert$ & $\vert {\rm Im}(\delta_{11}^{u})_{LR} \vert$ &
$\vert {\rm Im}(\delta_{22}^{d})_{LR} \vert$  &
$\vert {\rm Im}(\delta_{11}^{l})_{LR} \vert$ \\
\hline
\hline
0.3 & $4.3\times 10^{-8}$ & $4.3\times 10^{-8}$ & $3.6\times 10^{-6}$ & $4.2\times 10^{-7}$  \\
1 & $8.0\times 10^{-8}$ & $8.0\times 10^{-8}$ & $6.7\times 10^{-6}$ & $5.1\times 10^{-7}$\\
3 & $1.8\times 10^{-7}$ & $1.8\times 10^{-7}$ & $1.6\times 10^{-5}$ & $8.3\times 10^{-7}$ \\
\hline
\end{tabular}
\caption{MI bounds from the mercury EDM for an average 
squark mass of $600\mbox{ GeV}$ and for the electron EDM with
an average slepton mass of $320\mbox{ GeV}$ and different values of 
$x=m_{\tilde g}^2/m_{\tilde q}^2$. The bounds scale as 
$(m_{\tilde q (\tilde l)} (\mbox{GeV}) / 600 (320))$.}
\label{tab:MI2}
\end{center}
\end{table}
We compare these estimates with the phenomenological bounds in Table 
\ref{tab:MI2}\cite{hg,gabbiani}.
The bounds from neutron and electron EDM do not provide any new 
information on the structure of the Yukawa
textures.  However, the mercury EDM bounds are much more restrictive
and taking $x\simeq 1$ we find that, in the down sector, the case
$n=1$, $m=2$ is not allowed by EDM experiments
and we require $n \geq 2$, $m \geq 3$. As we said above, the same bound 
applyes for subdominant corrections to $Y_{2 2}$ and $Y_{1 2},Y_{2 1}$ 
where the first correction to the dominant terms can only be 
$\epsilon^4$ or $\epsilon^5$ respectively to satisfy EDM bounds.

In summary we have shown here that in a broken supergravity theory any field
that acquires a vev obtains at the same time an F-term of order 
$\langle \theta \rangle m_{3/2}$. These F-terms contribute unsuppressed
to the trilinear couplings and produce observable effects in the low energy 
phenomenology. We have seen that $\mu \to e \gamma$ decays can receive
large contributions from this source at the level of current experimental
bounds. Therefore this decay may provide the first clue of the structure
of the soft breaking sector with the proposed experiments in the near future.

We are grateful for useful conversations with R. Rattazzi.
We acknowledge support from the RTN European project HPRN-CT-2000-0148. O.V.
acknowledges partial support from the Spanish MCYT FPA2002-00612.

\end{document}